\documentclass[aps,prl,twocolumn,groupedaddress,showpacs,floatfix,superscriptaddress,longbibliography,10pt]{revtex4-2}
\usepackage{float}
\usepackage[utf8]{inputenc}
\usepackage[T1]{fontenc}
\usepackage{mathtools,amssymb,graphicx,units,bm}
\usepackage{soul} 
\usepackage[dvipsnames]{xcolor}
\usepackage[normalem]{ulem}
\usepackage{bbold} 
\usepackage{dcolumn}
\usepackage[version=4]{mhchem}
\usepackage{physics}%
\usepackage{silence}
\emergencystretch=\maxdimen
\usepackage[plainpages=false,pdfpagelabels,colorlinks=true,linkcolor=red,urlcolor=PineGreen,citecolor=PineGreen,pdftitle={Haldane_Hostein_QMC},pdfauthor={SJTSM},pdfdisplaydoctitle=true,pdfduplex=DuplexFlipLongEdge]{hyperref}
\usepackage{orcidlink}

\newcommand{\prlsection}[1]{\noindent\textit{#1.---}\hspace{0.25em}}
\newcommand{\beginsupplement}{%
        \setcounter{table}{0}
        \renewcommand{\thetable}{S\arabic{table}}%
        \setcounter{figure}{0}
        \renewcommand{\thefigure}{S\arabic{figure}}%
}

\def\s{\sigma}
\def\sp{\sigma^\prime}
\def\ave#1{\langle #1\rangle}
\newcommand{\avs}{\ave{\mathrm{sign}}}
\newcommand{\iv}{\mathbf{i}}
\newcommand{\jv}{\mathbf{j}}

\newcommand{\rv}{\mathbf{r}}
\def\rm#1{\mathrm{#1}}
\def\bf#1{\mathbf{#1}}

\begin{document}

\setstcolor{red}
\title {Real-space topology and charge order in the Haldane-Holstein Model}

\author{Sebasti\~ao dos A. Sousa-Júnior\,\orcidlink{0000-0002-4266-3780}}
\email{sebastiao.a.s.junior@gmail.com}
\affiliation{Department of Physics, University of Houston, Houston, Texas 77204, USA}
\author{Julián~Faúndez\,\orcidlink{0000-0002-6909-0417}}
\affiliation{Departamento de F\'isica y Astronom\'ia, Universidad Andr\'es Bello, Santiago 837-0136, Chile}
\author{Tarik~P.~Cysne\,\orcidlink{0000-0002-3830-3571}}
\affiliation{Instituto de Física, Universidade Federal Fluminense, Niterói, Av. Litorânea sn 24210-340, RJ-Brazil}
\author{Richard~T.~Scalettar\,\orcidlink{0000-0002-0521-3692}}
\affiliation{Department of Physics, University of California,
Davis, CA 95616, USA}
\author{Rubem~Mondaini\,\orcidlink{0000-0001-8005-2297}}
\affiliation{Department of Physics, University of Houston, Houston, Texas 77204, USA}
\affiliation{Texas Center for Superconductivity, University of Houston, Houston, Texas 77204, USA}

\begin{abstract}
We study the half-filled Haldane-Holstein model, where a paradigmatic Chern insulator is coupled to fully dynamical phonons, and provide an unbiased characterization of how retarded electron-phonon interactions destabilize Chern topology. Using determinant quantum Monte Carlo, we find that increasing the coupling drives an abrupt, first-order transition from a Chern insulator to a staggered charge-density wave that acts as a dynamical sublattice (Semenoff) mass. The transition is simultaneously signaled by a nearly quantized many-body Bott index and a real-space local Chern marker constructed from the interacting Green's function, both of which collapse as the charge order parameter becomes extensive. Spectral and open-boundary calculations reveal concomitant gap closing and the loss of boundary spectral weight at the critical coupling. Despite the generic phase problem induced by broken time-reversal symmetry, we show that it remains mild in the low-frequency regime and that the average phase factor sharply tracks the Chern insulator-charge density wave boundary. Our results establish a concrete route by which electron-phonon coupling can trigger a discontinuous collapse of Chern topology and provide experimentally relevant signatures for correlated topological platforms.
\end{abstract}

\maketitle


\prlsection{Introduction} Many topological phases of matter, such as quantum Hall insulators \cite{Thouless1982}, Chern insulators (CI) \cite{haldane1988}, and quantum spin Hall insulators \cite{Kane2005, Bernevig2006}, are traditionally described in terms of the band structure of non-interacting systems \cite{hasan2010}. The robustness of these phases to disturbances such as disorder and interactions has long been understood in weak-coupling regimes \cite{Laughlin1981, Xu2006}. However, the effects of strong electronic correlations on these topological phases have only recently been explored in depth \cite{Hohenadler2013, Rachel2018}. Electronic correlations can break the symmetries that protect topological phases \cite{Hohenadler2011, Hohenadler2012}, affect their characteristic edge states \cite{Varney2010, Zheng2011, Wu2012, Shao2021, He2024}, or even drive systems into exotic topologically ordered phases \cite{Maciejko2013, Ruegg2012, Grushin2015, Zhu2016, Mai2023}. The scrutiny of all possibilities arising from the interplay between electronic correlations and topology is just beginning. 

Among the different types of strongly correlated electronic states that may affect the topological properties of quantum materials, charge density waves (CDW) have attracted attention for appearing in a broad range of low-dimensional systems, including Dirac materials. In particular, substantial effort has been devoted to understanding CDW phases in quasi-two-dimensional transition-metal dichalcogenides (TMDs) and their competition or intertwining with superconductivity \cite{CastroNeto2001, Lian2018, Xi2015, Ugeda2015, Liu2021}. More recently, the interplay of translation-symmetry breaking and topological phases characterized by nonzero Chern number has been explored in bilayers, where the topology is experimentally evidenced by quantized Hall responses while CDW order is supported by commensurability/symmetry-breaking signatures \cite{Polshyn2022, Pan2022, Wilhelm2021}, and in layered three-dimensional materials exhibiting quantum Hall phenomenology \cite{Sasmal2022}. Microscopically, CDW instabilities can originate from Coulomb-driven electronic ordering or electron-phonon (\textit{e-ph}) coupling (or both), with direct real-space evidence for interaction-driven lattice-scale orders in graphene \cite{Coissard2022} and phonon-softening signatures of \textit{e-ph}-driven CDW formation in prototypical TMDs \cite{Weber2011}. A CDW-based mechanism has also been invoked to account for the field-induced 3D quantum Hall effect reported in ZrTe$_5$ \cite{Tang2019,Qin2020}.

The overall scenario described above underscores the need for an unbiased treatment of CDW-topology competition in the presence of retarded interactions. In this Letter, we address this problem in the Haldane-Holstein model \cite{Cangemi2019, Islam2024}, which combines the paradigmatic Chern insulator of Haldane \cite{haldane1988} with onsite Holstein phonon \textit{e-ph} coupling \cite{Holstein1959}. Employing large-scale determinant quantum Monte Carlo (DQMC) \cite{Blankenbecler1981,Hirsch1983,Hirsch1985,Scalettar1989,rrds2003,Assaad2008,Gubernatis16} and diagnosing topology via the many-body Bott index and local Chern marker \cite{Loring2010,Bianco2011,WeiChen2023}, we map the ground-state phase diagram and find that the CI gives way to a CDW phase through a first-order transition as one tunes the \textit{e-ph} coupling. While broken time-reversal symmetry generically introduces a phase problem in DQMC, we show that it is sufficiently mild in the low-frequency regime to permit reliable finite-size scaling, and that the average phase factor itself sharply tracks the CI-CDW phase boundary \cite{Mondaini2022,Mondaini2023}.


\prlsection{The Haldane-Holstein model} \label{sec:Method} The Hamiltonian is given as 

\begin{multline}
    \hat {\cal H} =   -t_{1} \sum_{\langle \iv,\jv \rangle, \s}
( \hat c^\dagger_{\iv\s} \hat c^{\phantom{\dagger}}_{\jv\s} + \mathrm{H.c.} )  +  t_{2} \sum_{\langle\langle \iv,\jv \rangle\rangle,\s}
( i\nu_{{\bf i}{\bf j}} \ \hat c^\dagger_{\iv\s} \hat c_{\jv\s}^{\phantom{\dagger}} + \mathrm{H.c.} )  \\ 
 + \sum_{\iv} \left[ \frac{\hat{P}_{\iv}^{2}}{2M} + \frac{M \omega_{0}^{2}}{2} \hat{X}^{2}_{\iv} \right]  - g \sum_{\iv} (\hat n_{\iv}-1) \hat{X}_{\iv} ,
\label{eq:Hamiltonian}
\end{multline}
where $\hat c_{\iv\sigma}^{\phantom{\dagger}}$ ($\hat c_{\iv\sigma}^\dagger$) annihilates (creates) an electron of spin $\sigma$ at site $\iv$ and
$\hat n_{\iv}=\hat c^\dagger_{\iv\uparrow}\hat c^{\phantom{\dagger}}_{\iv\uparrow}+\hat c^\dagger_{\iv\downarrow}\hat c^{\phantom{\dagger}}_{\iv\downarrow}$.
The first term describes nearest-neighbor hopping with amplitude $t_1$ in a honeycomb lattice. The second term is the complex next-nearest-neighbor hopping for the Haldane model, with amplitude $it_2$ and chirality factor $\nu_{\iv\jv}=\pm 1$. With our convention, $\nu_{\iv\jv}=\mathrm{sign}[({\bf d}_1\times{\bf d}_2)_z]=+1(-1)$ for hops on the $A(B)$ sublattice where ${\bf d}_1$ and ${\bf d}_2$ are the two bond directions traversed from $\jv$ to the next-nearest neighbor $\iv$. The third term describes dispersionless onsite phonons of frequency $\omega_0$ as quantum harmonic oscillators, and the last term is the onsite Holstein electron-phonon coupling of strength $g$.

\begin{figure}[t]
    \centering
    \includegraphics[scale = 0.5]{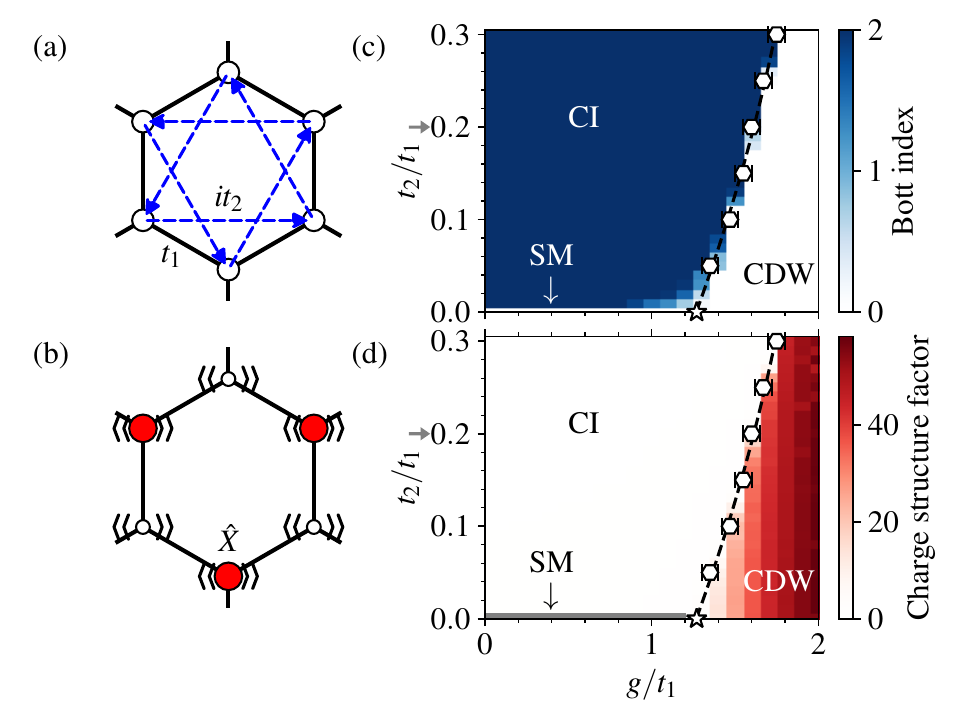}
    \caption{Schematic picture of the Haldane model hopping terms (a) and the charge-ordered phase (b) induced by Holstein phonons. Contour plots of the Bott index (c) and charge structure factor (d) are shown as functions of $g/t_1$ and $t_2/t_1$, for $L=6$, $\omega_0/t_1=1$, and $T/t_1=1/12$. The combination of both quantities allows us to map the semimetal (SM), charge density wave (CDW), and Chern insulator (CI) phases. Markers represent the critical values of $g$ obtained through finite-size scaling, and dashed lines serve as a guide; the star marker on the horizontal axis is from
    Ref.~\cite{Zhang20192}.
    Arrows at $t_2 =0.2t_1$ indicate the next-nearest-neighbor hopping value used for the phase diagram in Fig.~\ref{fig:diagram_omega}.
    }
    \label{fig:cr_cdw_map}
\end{figure}

Hereafter, we set $M=1$ and measure energies in units of $t_1$.
Unless stated otherwise, we fix $t_2=0.2t_1$ and impose periodic boundary conditions. All results are obtained at half-filling, where the competing Chern-insulating and CDW tendencies are most clearly exposed. A schematic representation of the model is shown in Figs.~\ref{fig:cr_cdw_map}(a) and \ref{fig:cr_cdw_map}(b)~\footnote{The choice $it_2$ corresponds to the Haldane phase $\phi=\pi/2$ \cite{haldane1988}.}.

We investigate Eq.~\eqref{eq:Hamiltonian} using finite-temperature determinant quantum Monte Carlo (DQMC) \cite{Blankenbecler1981,Gubernatis16,Hirsch1983,Hirsch1985,rrds2003,Assaad2008}. Because the Haldane term breaks time-reversal symmetry~\cite{haldane1988}, the fermionic weight is generically complex, and DQMC suffers from a phase (sign) problem. In the low-frequency phonon regime studied here, the latter is sufficiently mild to permit controlled finite-size scaling and an accurate determination of the CI-CDW critical points. In the antiadiabatic limit, $\omega_0\!\to\!\infty$, where the phase problem becomes severe, exact diagonalization (see Supplemental Materials~\cite{SM}) reveals enhanced onsite $s$-wave pairing coexisting with CDW correlations, consistent with the SU(2) pseudospin symmetry of the effective attractive-Hubbard description in this limit.

The emergence of charge order is quantified via equal-time correlations and their size dependence. In particular, we measure the staggered charge structure factor at $\mathbf{q}=0$,
\begin{equation}
S_{c} \equiv \frac{1}{N}\sum_{\mathbf{r},\mathbf{r}'} 
\left\langle \big(\hat n_{A,\mathbf{r}}-\hat n_{B,\mathbf{r}}\big)\big(\hat n_{A,\mathbf{r}'}-\hat n_{B,\mathbf{r}'}\big)\right\rangle ,
\label{eq:scdw}
\end{equation}
where $A,B$ label the two sublattices, $\mathbf{r}$ denotes unit-cell positions, and $N=2L^2$ is the total number of sites on an $L\times L$ honeycomb lattice. We additionally consider $m_c^2 \equiv S_c/N$, which extrapolates to the squared CDW order parameter as $L\to\infty$ to diagnose long-range staggered CDW order.

The topological transition is characterized using the local Chern marker $C_\sigma({\bf R})$ \cite{Bianco2011,WeiChen2023}, whose bulk spatial average approaches a quantized nonzero value in the topological phase. For noninteracting systems, it can be written as
\begin{align}
C_\sigma({\bf R})\!=\!4\pi\Im\bra{{\bf R}}\hat{\cal P}_\sigma\hat{R}_x \hat{\cal Q}_\sigma\hat{R}_y \hat{\cal P}_\sigma-\hat{\cal P}_\sigma\hat{R}_y \hat{\cal Q}_\sigma\hat{R}_x \hat{\cal P}_\sigma\ket{{\bf R}},
\label{Eq:marker}
\end{align}
where $\hat{\cal P}_\sigma$ and $\hat{\cal Q}_\sigma$ project onto occupied and unoccupied single-particle states of spin $\sigma$, respectively, and $\hat R_x,\hat R_y$ are site-position operators.
In DQMC, the projectors are not directly accessible; following Ref.~\cite{Melo2023}, we instead use the equal-time single-particle Green's function $\hat G^\sigma$ as an effective one-body density matrix, $\hat{\cal P}_\sigma\approx \hat G^\sigma$ and $\hat{\cal Q}_\sigma=\mathbb{1}-\hat G^\sigma$ (with the matrix elements $G^\sigma_{ij} = \langle \hat c_{i\sigma}^{\phantom{\dagger}}\hat c_{j\sigma}^\dagger\rangle$ obtained from DQMC). Although this procedure can suffer from reduced quantization near the transition due to finite-size effects and interaction-induced spectral broadening, it was shown in Ref.~\cite{Melo2023} to remain a reliable indicator of topological phases in interacting systems. Following Refs.~\cite{WeiChen2023,Melo2023}, we evaluate the Chern marker in the central (bulk) region of the lattice, $C\equiv\sum_\sigma C_\sigma({\bf R}_c)$, where ${\bf R}_c = {\bf a}_1L/2 + {\bf a}_2L/2$, and ${\bf a}_1$ and ${\bf a}_2$ are the primitive lattice vectors
\footnote{It is worth emphasizing that Green's-function-based topological markers have been successfully employed in QMC simulations of correlated phases~\cite{Hung2014,He2016,He2016_2}, but are known to suffer from reduced quantization near the topological phase transition. Other approaches to topological diagnostics in interacting systems were also previously explored~\cite{Hung2013,Gilardoni2025,Favata2025,Becca2025}.}.

As an alternative diagnostic, we also compute the Bott index \cite{Loring2010},
\begin{equation}
B=-\frac{1}{2\pi}\sum_\sigma\Im\,\mathrm{Tr}\left[\log\!\left(\hat V_\sigma^{\phantom{\dagger}}\hat W_\sigma^{\phantom{\dagger}}\hat V_\sigma^\dagger\hat W_\sigma^\dagger\right)\right],
\label{eq:bott}
\end{equation}
with $\hat W_\sigma=\hat{\cal P}_\sigma e^{i2\pi \hat R_x/L}\hat{\cal P}_\sigma+(\mathbb{1}-\hat{\cal P}_\sigma)$ and $\hat V_\sigma=\hat{\cal P}_\sigma e^{i2\pi \hat R_y/L}\hat{\cal P}_\sigma+(\mathbb{1}-\hat{\cal P_\sigma})$, where, as before, we construct $\hat{\cal P}_\sigma$ from $\hat G^\sigma$. The Bott index avoids gauge fixing and remains numerically stable for finite-size systems \cite{Loring2010}; it is also well quantized when a single-particle gap is present. For the spinful case studied here, the CI phase corresponds to $B=2$.

\begin{figure}[t]
    \centering
    \includegraphics[scale = 0.5]{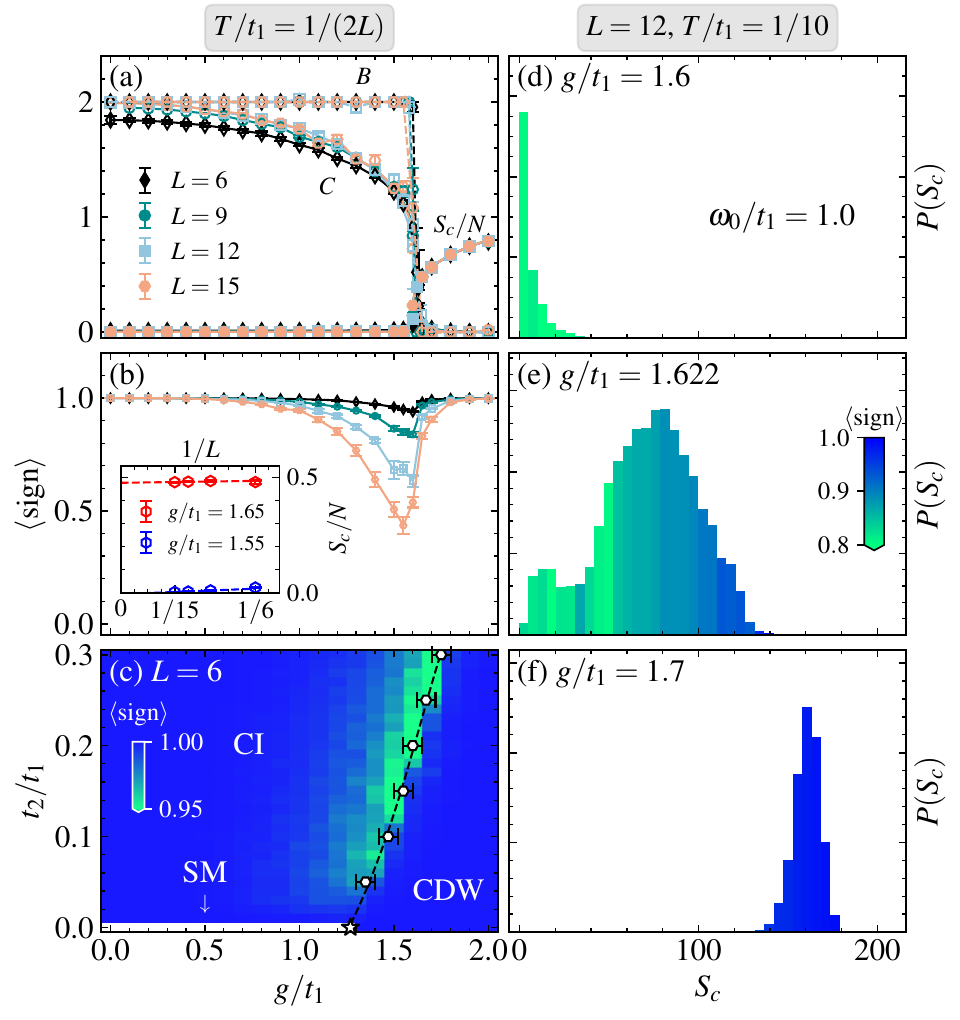}
    \caption{(a) Chern marker $C$, normalized charge structure factor and Bott index as functions of $g/t_{1}$. (b) Average determinant sign, $\avs $ versus $g/t_{1}$; inset: finite-size scaling of $S_c/N$ near the critical region. (c) Colormap of $\avs$ as a function of $g/t_{1}$ and $t_{2}/t_{1}$ for $L=6$ and $T/t_{1}=1/12$. (d)–(f) Reweighted histograms of $S_{c}$ at $g/t_{1}=1.6$, $1.622$, and $1.7$, respectively, computed for $L=12$ and $T/t_{1}=1/10$; the color of each histogram bar encodes the average sign for that reweighted bin. All data for $\omega_0/t_1 = 1$.
    }
    \label{fig:cr_cdw_cut}
\end{figure}

\prlsection{\text{CI-CDW} transition}
\label{subsec:CDW_transition}
Starting from the $g = 0$ limit, i.e., no phonon-coupling, the system is known to be a Chern insulator for any $|t_2/t_1| > 0$~\cite{haldane1988}. In contrast, for $t_2 = 0$, a semimetal (SM)-to-CDW transition occurs at $g/t_1 \approx 1.27$~\cite{Zhang20192,Chen2019}. When both $g$ and $t_2$ are finite, a competition arises between topology and charge order. Figure\,\ref{fig:cr_cdw_map} summarizes this ground-state behavior, showing maps of the Bott index [Fig.\,\ref{fig:cr_cdw_map}(c)] and the charge structure factor [Fig.\,\ref{fig:cr_cdw_map}(d)] in the $g\times t_2$ plane for $L = 6$, $T/t_1 = 1/12$ and $\omega_0/t_1=1$. The Bott index remains nearly quantized ($B\simeq 2$) at small $g/t_{1}$ and finite $t_{2}/t_{1}$, signaling a stable topological phase, but vanishes at large $g/t_{1}$. In contrast, $S_{c}$ exhibits a strong enhancement in this latter regime. Our results thus indicate that the onset of CDW order coincides with the disappearance of the topological phase. Indeed, the transition lines extracted from the two quantities agree closely with those indicated by the markers, obtained from finite-size scaling as discussed below.

In line with the previous results, Fig.\,\ref{fig:cr_cdw_cut}(a) shows the Chern marker, the Bott index, and the normalized charge structure factor $S_{c}/N$ as functions of $g/t_{1}$ for larger system sizes, at temperatures $T/t_1 = 1/(2L)$. While the Chern marker is finite for $g/t_{1} \lesssim 1.6$, it vanishes for larger values of $g/t_{1}$, coinciding with a sharp drop of the Bott index from finite to zero. Conversely, $S_c$ turns dominant for $g/t_{1} \gtrsim 1.6$, signaling the emergence of long-range charge order. The abrupt increase of $S_{c}/N$ near $g/t_{1}\!\sim\!1.6$ is suggestive of a first-order transition into the CDW phase. For even larger lattice sizes and lower temperatures, finite-size effects are expected to diminish, leading to a Chern marker that approaches a quantized value near the transition and vanishes deep in the CDW regime. The Bott index, in turn, remains well quantized except near the critical region, where the gap closing prevents exact quantization. 

One can physically understand the competition between the CDW and CI phases by recognizing that the CDW order corresponds to a spontaneous breaking of sublattice symmetry present in the Hamiltonian of Eq.\,\eqref{eq:Hamiltonian} (see SM), induced by electron-phonon interaction, effectively acting as a Semenoff mass \cite{Bernevig20062}. It is well known that sublattice symmetry-breaking, even if not spontaneous as in the non-interacting Haldane model~\cite{haldane1988}, can destroy the non-trivial topology of the system.

\prlsection{Sign problem}
\label{subsec:signproblem}
Interestingly, the CI-CDW transition is also reflected in the behavior of the average determinant sign of the fermionic $\mathcal{M}$-matrices in DQMC, $\avs \equiv |\langle e^{i\theta} \rangle|$, where $\theta = \arg(\det[\mathcal{M}^\uparrow \mathcal{M}^\downarrow])$. Figure\,\ref{fig:cr_cdw_cut}(b) shows $\langle \mathrm{sign} \rangle$ as a function of $g/t_1$. Although the sign decreases with increasing lattice size, all curves display a pronounced minimum near the critical point $g/t_1\!\sim\!1.6$. Finite-size scaling of the charge structure factor further confirms the absence (presence) of charge order for $g/t_1\!=\!1.55$ ($g/t_1\!=\!1.65$), as shown in the inset of panel (b). Moreover, the overall behavior of $\langle \mathrm{sign} \rangle$ serves as a useful proxy of the phase boundary~\cite{Mondaini2022,Mondaini2023,Yi2024} across almost the entire $g \times t_2$ parameter space, as illustrated in Fig.\,\ref{fig:cr_cdw_cut}(c), where a dip in the sign separates the CI and CDW phases. Finally, the critical values of $g$ (black markers) and the minimum of $\langle \mathrm{sign} \rangle$ converge to the critical point obtained for the topologically trivial ($t_2\!=\!0$) case~\cite{Zhang20192}, indicated by the white star in Fig.\,\ref{fig:cr_cdw_cut}(c).

Further clarification of the transition order and the behavior of $\avs$ across the critical region is obtained by histograms of $S_c$ near the transition, shown in Fig.\,\ref{fig:cr_cdw_cut}(d)-(f). Fixing $L = 12$ and $T/t_1= 1/10$, and varying the \textit{e-ph} coupling within a narrow window $\delta g = 0.1t_1$, we observe pronounced changes in the mean values, shapes, and sign distributions as the transition is crossed. For $g/t_1 = 1.6$, Fig.~\ref{fig:cr_cdw_cut}(d) exhibits a single-peaked distribution centered at small $S_c$, while for $g/t_1 = 1.7$ [Fig.\,\ref{fig:cr_cdw_cut}(f)] the histogram shifts sharply to larger values. At intermediate coupling, $g/t_1=1.622$ [Fig.\,\ref{fig:cr_cdw_cut}(e)], the distribution becomes clearly bimodal, consistent with phase coexistence at a first-order transition \cite{Schmid2004,Isakov2006}. The average determinant sign offers an additional visualization of the distinction between the two phases: within each bin, $\avs$ assumes higher (lower) values in regions with stronger (weaker) charge correlations. A comparable evolution of $\avs$ has been reported at the CDW transition in the extended Hubbard model~\cite{Sousa-Junior2024,Kennedy2025}.


\prlsection{Spectral Properties}
\label{subsec:Spectral}
We now study how the \textit{e-ph} and the CI-CDW transition renormalize single-particle excitations. For that, we compute the sublattice-resolved spectral function $A_{\gamma}(\mathbf{k},\omega)$ ($\gamma=A,B$) from the corresponding imaginary-time Green's functions $G_{\gamma}(\mathbf{k},\tau)$ obtained in DQMC (see \cite{SM}). These are related by the fermionic spectral representation $G_{\gamma}(\bf k,\tau) =  \int d{\omega} K(\tau,\omega) A_{\gamma}({\bf k},\omega)$ \cite{Silver1990,beach2004}, with kernel $K(\tau,\omega)=\frac{e^{-\omega \tau}}{1+e^{-\omega/T}}$, from which $A_{\gamma}({\bf k},\omega)$ is reconstructed by performing Maximum Entropy analytic continuation \cite{Assaad2022}.

\begin{figure}[t]
    \centering
    \includegraphics[scale = 0.53]{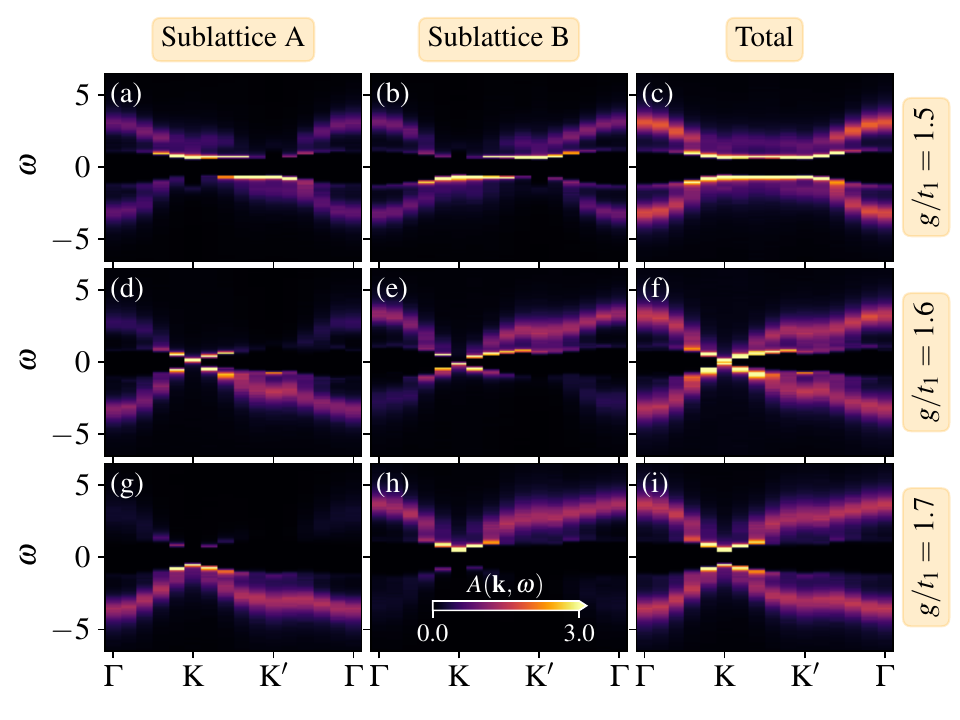}  
    \caption{Single-particle spectral function for $L=15$, $t_2/t_1=0.2$, and $T/t_1=1/30$. (a)-(c) show the CI phase at $g/t_1=1.5$: spectral weight resolved on sublattice $A$ [(a)] and $B$ [(b)], and the total spectral function [(c)]. (d)–(f) correspond to the critical coupling $g/t_1=1.6$, and (g)–(i) show the CDW phase ($g/t_1=1.7$).}
    \label{fig:spectral}
\end{figure}

Figure~\ref{fig:spectral} shows the single-particle spectral function across the CI-CDW transition. In the CI phase at $g/t_1=1.5$, the spectrum exhibits a reversal of low-energy sublattice-resolved spectral weight near the valleys, consistent with a band-inversion picture [Figs.~\ref{fig:spectral}(a,b)], while the total spectral function remains gapped [Fig.~\ref{fig:spectral}(c)]. At the critical coupling $g/t_1=1.6$, the spectra show both the persistence of valley-dependent features and an incipient charge imbalance: the $A$-sublattice (defined here as the one that becomes higher-density in the CDW phase) gains weight below the Fermi level, whereas the $B$-sublattice gains weight above it [Figs.~\ref{fig:spectral}(d,e)]. Correspondingly, the low-energy gap is strongly suppressed and appears to close near the $K$ valley before reopening for $g/t_1>1.6$ [Fig.~\ref{fig:spectral}(f)], consistent with a mass-inversion scenario at the topological transition~\footnote{Note that $K$ and $K'$ need not be equivalent because time-reversal symmetry is broken}. Deep in the CDW phase at $g/t_1=1.7$, the sublattice polarization becomes pronounced: the higher-density sublattice carries spectral weight predominantly below the Fermi level, and the lower-density sublattice above it [Figs.~\ref{fig:spectral}(g,h)], and the total spectrum displays a clear CDW gap [Fig.~\ref{fig:spectral}(i)].

While finite-size effects and limitations of the analytic continuation can smear spectral features in the vicinity of the critical point, the single-particle gap remains robust outside this region. This gap, in turn, underlies the observed quantization of the Bott index. In the SM\,\cite{SM}, we provide further analysis of the charge imbalance through the CI-CDW transition.

\begin{figure}[t]
    \centering
    \includegraphics[scale = 0.5]{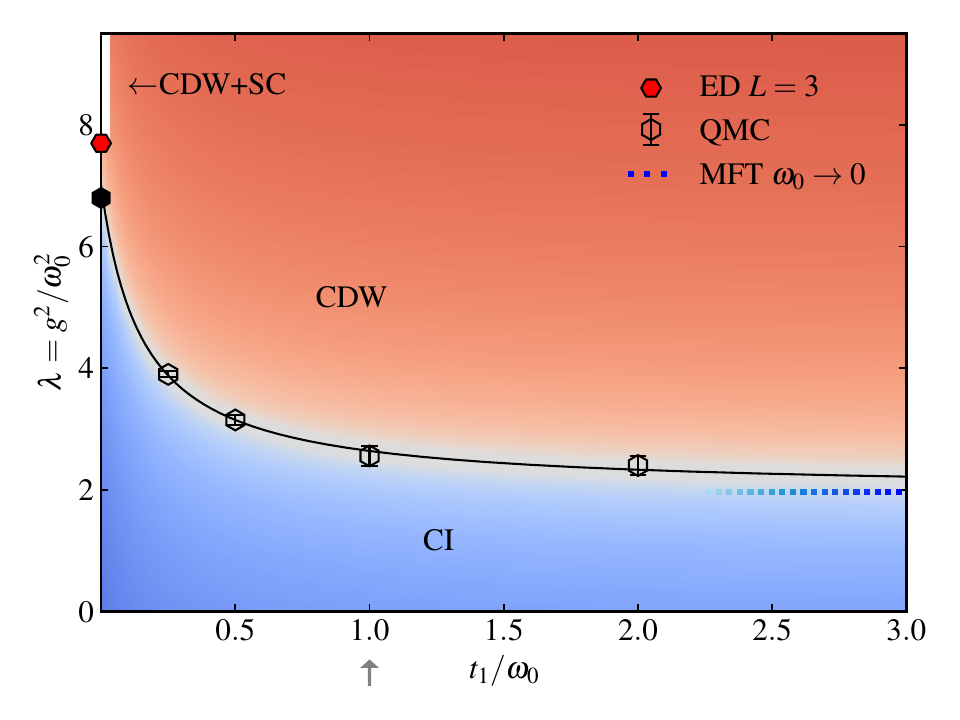}
    \caption{Phase diagram of the Haldane-Holstein model for $t_{2}/t_{1} = 0.2$. The Chern insulator (CI) phase appears at small values of $\lambda\! =\!g^{2}/\omega_{0}^{2}$, while the charge-density-wave (CDW) phase emerges at stronger couplings. In the $\omega_{0} \!\to\! \infty$ limit, superconductivity (SC) coexists with CDW, as confirmed by exact diagonalization (red marker). The solid black marker denotes the critical point obtained from density matrix renormalization group (DMRG) calculations~\cite{He2024}. Open black markers represent the determinant quantum Monte Carlo (DQMC) critical points, and the solid line is a guide to the eye. The blue dotted line indicates the mean-field critical $\lambda$, which becomes exact in the $\omega_{0} \!\to\! 0$ limit.  
    The arrow at $\omega_0 = t_1$ indicates the phonon frequency used in the phase diagram of Fig.~\ref{fig:cr_cdw_map}.
    }
    \label{fig:diagram_omega}
\end{figure}

\prlsection{Phonon frequency effects}
\label{sec:phonon-freq}
Having established the low-temperature phase structure at $\omega_0=t_1$, we now examine how the phonon frequency reshapes the CI-CDW boundary by tuning retardation. Figure~\ref{fig:diagram_omega} summarizes the phase diagram of Eq.~\eqref{eq:Hamiltonian} as a function of $t_{1}/\omega_{0}$ and $\lambda\!=\!g^{2}/\omega^{2}_{0}$, which sets the strength of the effective static attractive interaction. For weak coupling, the CI phase persists, while increasing $\lambda$ drives a transition into the CDW phase (open markers: DQMC critical points). 

In the antiadiabatic limit ($\omega_0\!\to\!\infty$), charge order coexists with superconductivity (SC), as confirmed by exact diagonalization (see~\cite{SM}). This coexistence follows from the SU(2) symmetry of the effective attractive Hubbard term, $U/t_1\equiv-\lambda$, which preserves degeneracy between charge and pairing channels, in contrast with the Kane-Mele-Hubbard model~\cite{Zheng2011,Hohenadler2011}. Through a particle-hole transformation, the Haldane-Hubbard model maps the attractive and repulsive cases, allowing direct comparison with density matrix renormalization group results that place the critical interaction at $|U_c|/t_1\!\sim\!6.8$~\cite{He2024}. In the opposite adiabatic limit ($\omega_{0}\!\to\!0$), the mean-field prediction $\lambda_{c}\!\sim\!2$ (blue dotted line) becomes asymptotically exact since lattice fluctuations are quenched.

\begin{figure}[t]
    \centering
    \includegraphics[scale = 0.5]{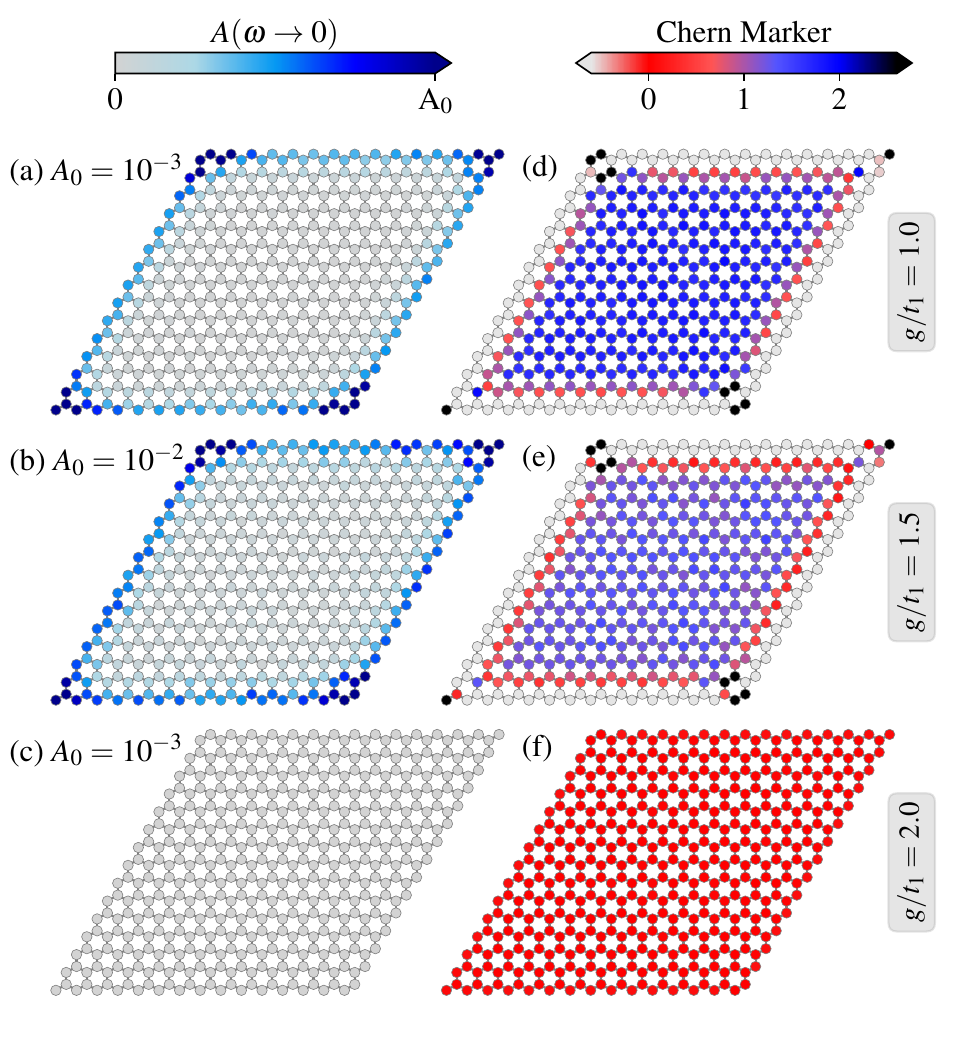}
    \caption{Results for open boundary conditions. Local density of states near the Fermi level for (a) $g/t_{1}=1.0$, (b) $g/t_{1}=1.5$, and (c) $g/t_{1}=2.0$. Panels (d)–(f) show the corresponding real-space local Chern marker $C(\mathbf{R})$. Data are for $L=15$, $\omega_{0}/t_{1}=1$, $t_{2}/t_{1}=0.2$, and $T/t_{1}=1/30$.}
    \label{fig:edge_modes}
\end{figure}

\prlsection{Open boundary conditions}
\label{sec:OBC}
To further probe the fate of topology deep in the interacting regime, we analyze open boundary conditions (OBC) and compute a proxy for the local density of states (LDOS) at the Fermi level. The low-energy spectral weight can be estimated directly from the imaginary-time Green's function via $A(\omega\!\to\!0)\simeq (1/\pi T)\,G(\tau=1/2T)$, which corresponds to the density of states thermally averaged over an energy window of width $\sim T$~\cite{Trivedi1995}.

Figure~\ref{fig:edge_modes} summarizes the OBC results. For $g/t_1=1$ [Fig.~\ref{fig:edge_modes}(a)], the LDOS near $\omega=0$ is strongly enhanced at the sample boundaries, consistent with chiral edge states, while the bulk topological character is captured by a nonzero local Chern marker in the central region [Fig.~\ref{fig:edge_modes}(d)]. Closer to the transition, at $g/t_1=1.5$, the boundary enhancement is reduced but remains visible [Fig.~\ref{fig:edge_modes}(b)], and although the Chern marker shows degraded quantization due to finite-size/finite-temperature effects, its finite bulk value still indicates a topological phase [Fig.~\ref{fig:edge_modes}(e)], in line with Ref.~\cite{Melo2023}. Deep in the CDW phase at $g/t_1=2$, both the boundary LDOS enhancement and the bulk Chern marker vanish [Figs.~\ref{fig:edge_modes}(c,f)], indicating the destruction of the Chern-insulating state.


\prlsection{Summary and outlook}
\label{sec:Conclusions}
Using unbiased DQMC, we map the low-temperature phase diagram of the Haldane-Holstein model versus phonon coupling and frequency. Topological markers (local Chern marker and Bott index) together with charge correlations and phase-reweighted histograms reveal a robust Chern insulator at weak coupling that collapses abruptly into a staggered CDW at strong coupling. For $t_2/t_1=0.2$ and $\omega_0/t_1=1$, the transition occurs at $g/t_1\simeq 1.6$ and shows clear first-order signatures. Spectral and OBC results are consistent with gap closing/reopening and the loss of boundary spectral weight across the transition. In the antiadiabatic limit, ED yields the expected attractive-Hubbard behavior with degenerate charge and onsite pairing correlations. These findings identify electron-phonon coupling as a route to an abrupt destruction of Chern topology via CDW formation, with direct experimental consequences for spectroscopies, scanning tunneling microscopy, and diffraction in candidate platforms including layered quantum Hall insulators \cite{Qin2020,Tang2019} and TMD mono- and heterobilayers \cite{Pan2022,Polshyn2022,Wilhelm2021,Zheng2019,Lin2020,Wang2023}.

S.A.S.-J. gratefully acknowledges financial support from the Brazilian Agency CNPq. J. F. acknowledges support from ANID Fondecyt grant number 3240320. TPC acknowledges CNPq (Grant No. 305647/2024-5). Powered@NLHPC: This research was partially supported by the NLHPC supercomputing infrastructure (CCSS210001). Part of the calculations used resources from the Research Computing Data Core at the University of Houston. This work also used TAMU ACES at Texas A\&M HPRC through allocation PHY240046 from the Advanced Cyberinfrastructure Coordination Ecosystem: Services \& Support (ACCESS) program, which is supported by U.S. National Science Foundation grants 2138259, 2138286, 2138307, 2137603, and 2138296. R.M.~acknowledges support from the T$_{\rm c}$SUH Welch Professorship Award.    RTS is supported by the grant DOE DE-SC0014671, funded by
the U.S.~Department of Energy, Office of Science. The data that support
the findings of this article are openly available at \cite{SasJr2026}.

\color{black}

\bibliography{references}

\beginsupplement

\clearpage

\setcounter{equation}{0}
\setcounter{enumiv}{0} 

\onecolumngrid
\begin{center}
\phantomsection
\label{sec:sm}
{\large\textbf{SUPPLEMENTAL MATERIAL}} \\
\textbf{Real-space topology and Charge Order in the Haldane-Holstein Model}\\
\end{center}
\onecolumngrid

These Supplemental Materials provide additional results supporting the main text. We present additional data on the charge structure factor and the associated correlation ratio, and use them to quantify how the Haldane hopping $t_2/t_1$ and the phonon frequency $\omega_0$ reshape the CDW transition. We also report the total energy and its electronic and phononic contributions across the transition. In addition, we study the effect of explicitly breaking sublattice symmetry by adding a staggered sublattice potential. Finally, we analyze the adiabatic ($\omega_0\!\to\!0$) and antiadiabatic ($\omega_0\!\to\!\infty$) limits using mean-field theory and exact diagonalization, respectively.

\section{Additional data}

\begin{figure}[H]
    \centering
    \includegraphics[scale=0.5]{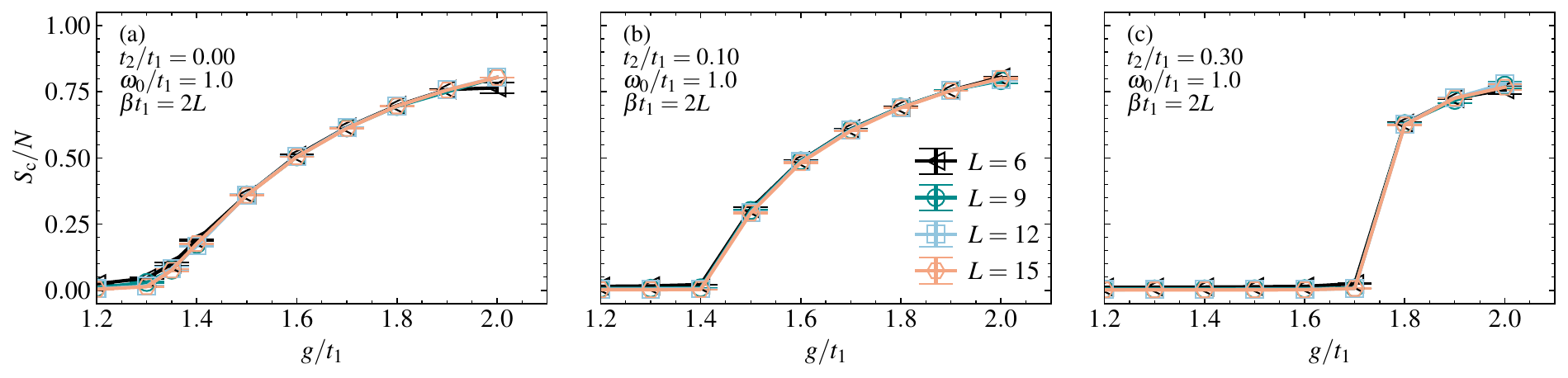}
    \caption{Scaled charge-density-wave (CDW) structure factor $S_c/N$ versus electron–phonon coupling $g/t_1$ for several lattice sizes $L$ and next-nearest-neighbor hopping ratios $t_2/t_1$, with phonon frequency fixed at $\omega_0/t_1=1.0$. (a)–(c) show $t_2/t_1=0.0$, $0.1$, and $0.3$, respectively. The inverse temperature is set to $\beta t_1 =2L$.}
    \label{fig:S1}
\end{figure}

In the main text, we have presented results of the scaled charge density wave (CDW) structure factor $S_{c}/N$ considering $t_{2}/t_{1}=0.2$, as shown in Fig.~\ref{fig:cr_cdw_cut}(a). To further elucidate the role of the next-nearest-neighbor hopping $t_{2}/t_{1}$, Fig.~\ref{fig:S1} displays its effects on the evolution of $S_{c}/N$ as a function of electron-phonon coupling $g$ for different lattice sizes $L$ in the low-temperature regime, for phonon frequency $\omega_{0}/t_{1}=1$. In general, the stabilization of the CDW order shows a pronounced dependence on $t_{2}/t_{1}$, as the onset of charge ordering occurs at larger critical values of $g/t_{1}$. But most prominently, the increase in charge correlations becomes abrupt for finite $t_2/t_1$, indicating a change in the transition character from second to first order.

\begin{figure}[t]
    \centering
    \includegraphics[scale = 0.5]{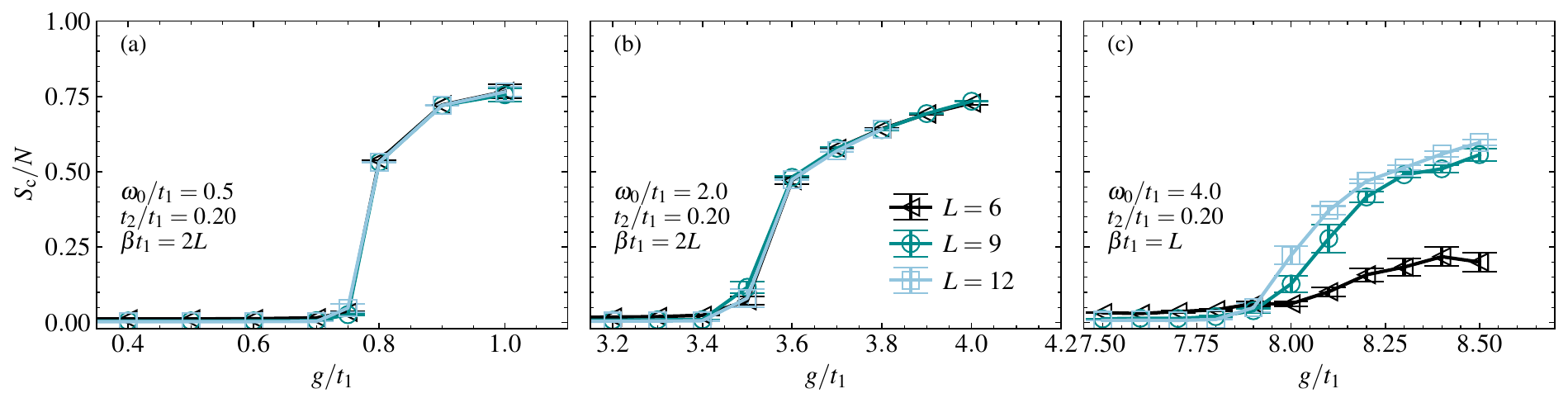}
    \caption{The scaled change-density-wave (CDW) structure factor as a function of electron-phonon coupling $g/t_{1}$ is shown for different values of lattice sizes $L$ and phonon frequency $\omega_{0}/t_{1}$, with the next-nearest-neighbor hopping radio fixed in $t_{2}/t_{1}=0.2$. Panels (a)-(c) correspond to $\omega_{0}/t_{1}=0.5$, 1.0, and 4.0, respectively.
    }
    \label{fig:S2}
\end{figure}

In contrast, Fig.~\ref{fig:S2} illustrates the evolution of $S_{c}/N$ as a function of $g$ for different $\omega_{0}$, while keeping $t_{2}/t_{1}$ fixed. These results reveal that $\omega_{0}$ plays a critical role in determining the onset of charge ordering, as the critical coupling $g$ shifts systematically with $\omega_{0}$.  Increasing the phonon frequency weakens the CDW order, and the enhancement of CDW correlations becomes much less pronounced. At zero temperature, however, we argue that the transition should remain first order in the high-frequency regime, consistent with exact-diagonalization (ED) results in the antiadiabatic limit discussed below. Note that, due to the sign (phase) problem, we relaxed the relation down to $\beta t_1 = L$ for $\omega_0/t_1=4$. Due to a more severe sign problem, determinant quantum Monte Carlo (DQMC) simulations for $\omega_0/t_1 >4$ are unfeasible for $t_2/t_1=0.2$.

\begin{figure}[H]
    \centering
    \includegraphics[scale = 0.5]{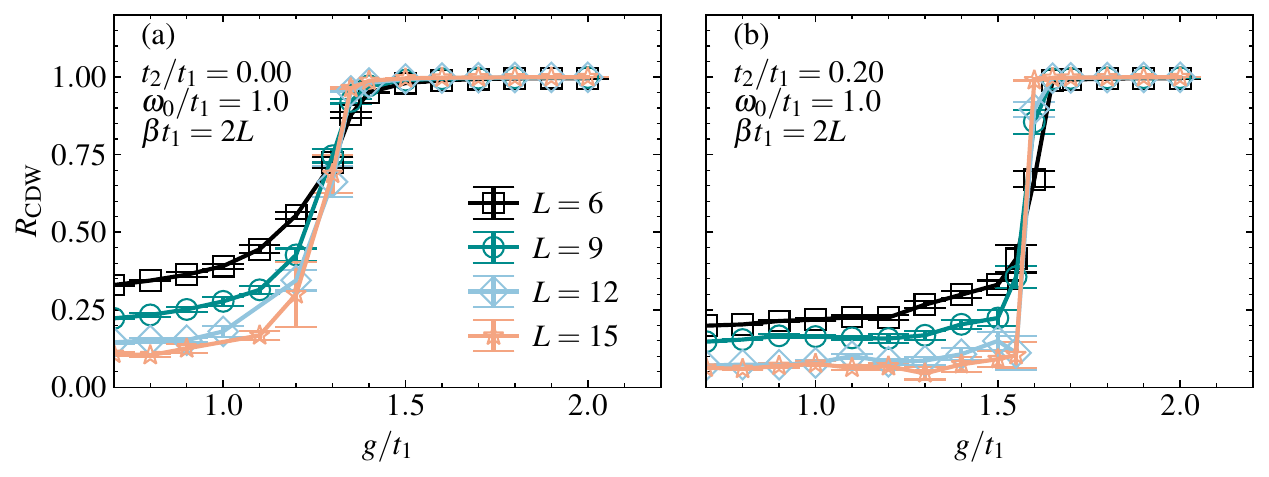}
    \caption{The charge-density-wave (CDW) correlation ratio $R_{\rm {CDW}}$ as a function of electron-photon coupling $g/t_{1}$ under different lattice sizes $L$ and next-nearest-neighbor hopping ratio $t_{2}/t_{1}$, with the phonon frequency fixed in $\omega_{0}/t_{1}=1.0$. In panel (a) $t_{2}/t_{1}=0.0$ and in (b)$t_{2}/t_{1}=0.2$, respectively.  The inverse of temperature is set to $\beta t_1 =2L$.}
    \label{fig:Supp_Rc}
\end{figure}

As an additional proxy of criticality, Fig.~\ref{fig:Supp_Rc} shows the CDW correlation ratio $R_{\rm {CDW}}$ as a function of $g/t_1$ for several lattice sizes $L$ at fixed $\omega_0/t_1=1$. We define
\begin{equation}
    R_{\rm {CDW}}=1-\frac{S_{c}(\mathbf{q}+\delta\mathbf{q})}{S_{c}(\mathbf{q})},
    \tag{S1}
\end{equation}
where $\mathbf{q}=(0,0)$ is the ordering wave vector of the staggered CDW on the honeycomb lattice and $\delta\mathbf{q}$ is one of the two inequivalent nearest-neighbor momenta to $\mathbf{q}$. In continuous transitions, $R_{\rm {CDW}}$ is approximately size independent at criticality, leading to a crossing of the curves for different $L$ near $g_c/t_1$. The observed crossings are consistent with the phase boundaries inferred from $S_c/N$ scaling [Figs.~\ref{fig:S1} and \ref{fig:cr_cdw_cut}(a)] and with the CI-CDW boundary in Fig.~\ref{fig:diagram_omega}. For $t_2/t_1=0.2$, the rapid evolution of $R_{\rm {CDW}}$ across the crossing, together with the abrupt behavior of $S_c/N$ and the bimodal histograms in the main text, supports a first-order CI--CDW transition.

\begin{figure}[H]
    \centering
    \includegraphics[scale= 0.5]{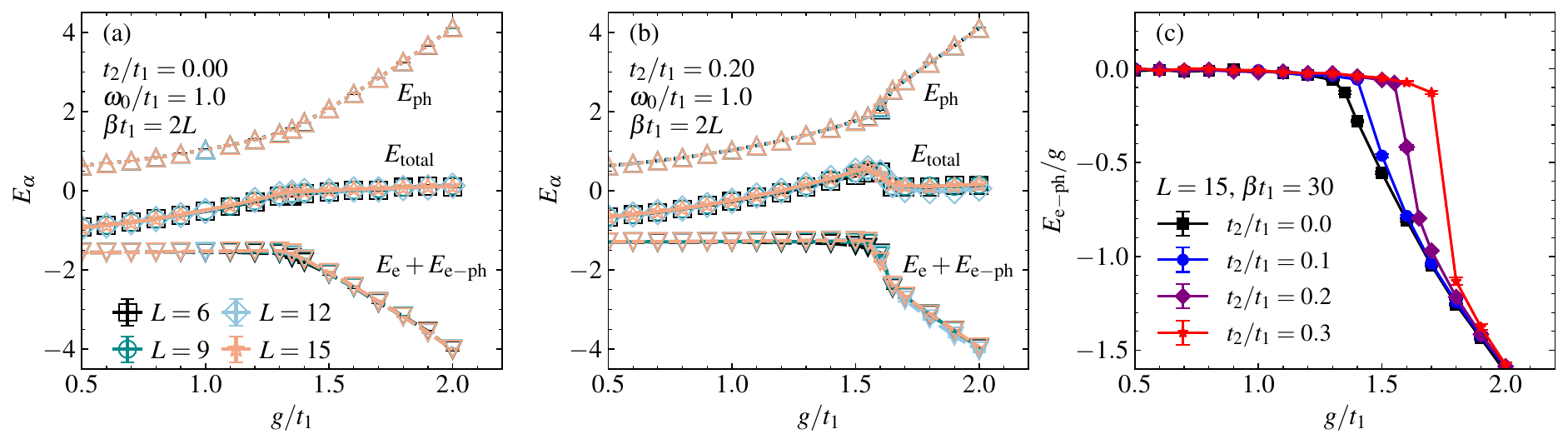}
    \caption{The average energy $E_{\alpha}$ per site as a function of electron-photon coupling $g/t_{1}$ under different lattice sizes $L$ and next-nearest-neighbor hopping ratio $t_{2}/t_{1}$, with the phonon frequency fixed in $\omega_{0}/t_{1}=1.0$. The symbol $\alpha$ represents the average energy associated with phonons, the combined electron-phonons, and the averaged total energy. In panel (a) $t_{2}/t_{1}=0.0$ and in (b)$t_{2}/t_{1}=0.2$, respectively.  The inverse of temperature is set to $\beta t_1=2L$. (c) Normalized $e-ph$ energy $E_{\rm e-ph}/g$ with respect to $g$, shown for several values of $t_{2}$. } 
    \label{fig:penergy}
\end{figure}

Figure~\ref{fig:penergy} shows the average energy components $E_{\alpha}$ as functions of $g/t_{1}$ for different lattice sizes $L$ and values of $t_{2}/t_{1}$. The presented quantities correspond to the phonon energy contribution ($E_{\text{ph}}$), the sum of the electronic kinetic energy ($E_\text{e}$) and the electron–phonon interaction energy ($E_{e+\text{ph}}$), and the total energy ($E_{\rm total}$). While in Fig.\,\ref{fig:penergy}(a), a change of behavior is observed near $g_c/t_1\approx1.3$ for $t_2/t_1=0.0$, a pronounced drop is present in the energy curves for $E_{\rm {total}}$ and $E_{\rm e} + E_{\rm {e-ph}}$ in panel (b), near the critical coupling $g_c/t_1\approx1.6$ for $t_2/t_1=0.2$, suggesting the first-order character of the transition. The sharp drop in the total energy at the critical region for $t_2>0$ is mainly associated with the term $E_{\text{e-ph}}=-\frac{g}{N}\sum_{\mathbf{i}}\langle (n_{\mathbf{i}}-1)\hat{X}_{\mathbf{i}}\rangle$. To emphasize this feature, we also plot $E_{\rm e-ph}/g$ as a function of $g/t_{1}$. As $t_{2}$ increases, this quantity evolves into an almost step-like function.

\section{Effects of a static staggered potential}

\begin{figure}[H]
    \centering
    \includegraphics[scale = 0.5]{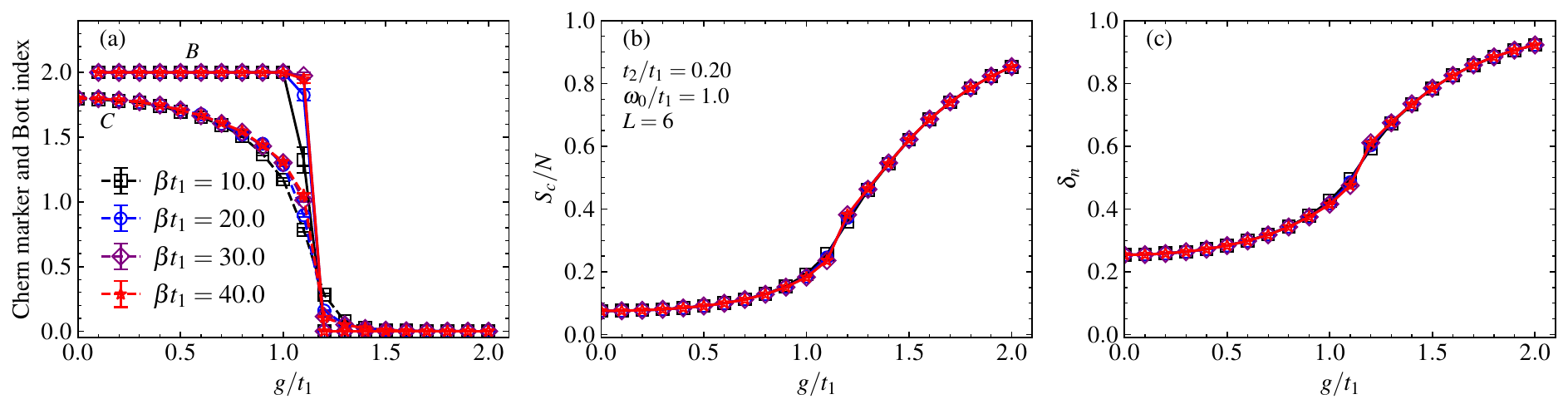}
    \caption{(a) The Chern marker ($C$) and the Bott index ($B$), (b) the change-density-wave (CDW) structure factor $S_{c}/N$,  and (c) the static staggered potential ($\delta_{n}$) as a function of electron-phonon coupling $g/t_{1}$ under different values of inverse of temperature  $\beta$ and fixed $\Delta/t_1 = 0.5$, next-nearest-neighbor hopping ratio $t_{2}/t_{1}=0.2$, phonon frequency $\omega_{0}/t_{1}=1$ and lattice size $L=6$. 
    }
    \label{fig:stag}
\end{figure}

As an additional result, in Fig.\,\ref{fig:stag}, we examine the effects of introducing a static sublattice-staggered potential term, $- \Delta \sum_\iv \tau_\iv n_\iv$~\cite{haldane1988}, where $\tau_i = +1$ for sublattice A and $\tau_i = -1$ for sublattice B. Panels (a)–(c) show the evolution of the Chern ($C$) and Bott ($B$) indices,  $S_{c}/N$, and the sublattice charge imbalance, $\delta_n = (\langle \hat n_A \rangle - \langle \hat n_B \rangle)/2$, as functions of $g/t_{1}$ for various inverse temperatures $\beta$. The inclusion of the staggered potential reduces the lattice symmetry to $C_{3v}$ and energetically favors electron occupation on one sublattice over the other, thus explicitly generating a CDW pattern.

As such, for $\Delta > 0$, a finite charge imbalance appears for any value of $g/t_{1}$ as observed in Fig.\,\ref{fig:stag}(c). However, since $\Delta < 3\sqrt{3}t_2$ and $|t_2/t_1 < 1/3|$~\cite{haldane1988}, a topological phase persists at low temperatures for $g/t_1 \lesssim 1.1$, characterized by a quantized, finite value of $B$. Figure~\ref{fig:stag} shows that for $g/t_1 \gtrsim 1.1$, both $C$ and $B$ vanish as the charge imbalance is enhanced by the charge correlations promoted by $g/t_{1}$ interaction. The system therefore exhibits coexistence of (explicitly broken) CDW and CI phases only when $|\Delta| > 0$. It is worth noting that the Bott index and the Chern marker probe only the topological transition, which in the present case does not necessarily correspond to the CDW transition; the latter, in turn, can only be probed by the charge correlations. 

\section{The Static Mean-Field Approximation}

\begin{figure}[H]
    \centering
    \includegraphics[scale = 0.5]{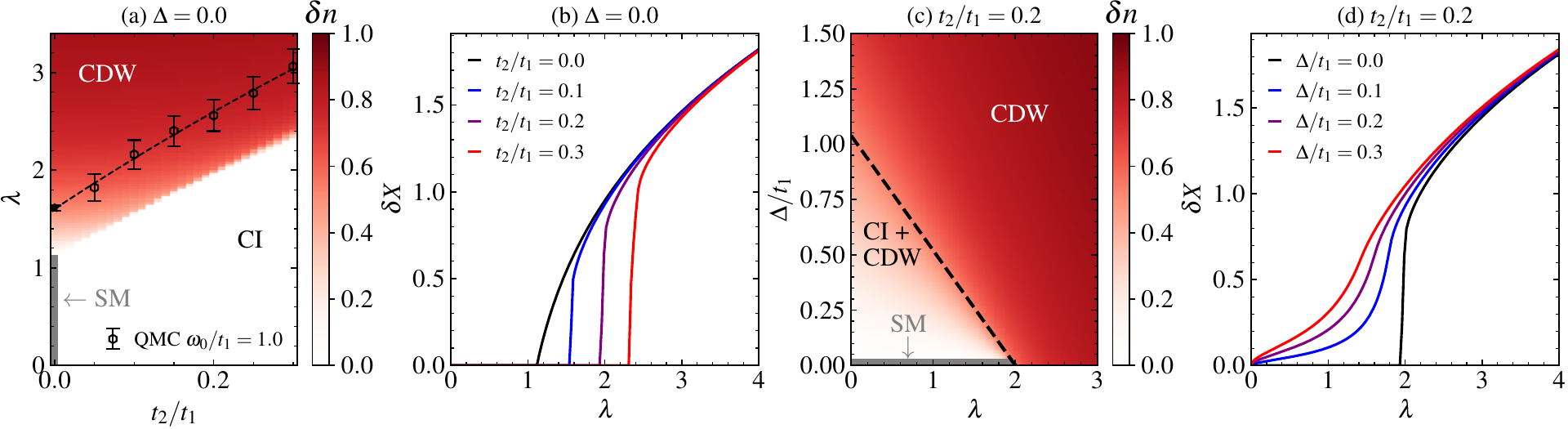}
    \caption{Mean-field results: (a) $t_2 \times \lambda$ ground state phase diagram and (b) line cuts of the phonon displacement $\delta X$ as a function of $\lambda$ for several values of $t_2/t_1$; (c) $\lambda \times \Delta$ ground state phase diagram and (d) line cuts of the phonon displacement $\delta X$ as a function of $\lambda$ for several values of $\Delta/t_1$. The red color in panels (a) and (c) maps the intensity of the charge imbalance $\delta_n$.     Black markers in panel (a) display the QMC critical points $\lambda_c = g_c^2/\omega_0^2$ for $\omega_0/t_1=1.0$.
    }
    \label{fig:mft}
\end{figure}

In the static mean-field approximation, we replace the phonon operator by its expectation value using the ansatz $\hat{X}_\iv \;\rightarrow\; X_0 +\tau_\iv \delta X $. We consider the adiabatic limit $\omega_0 \to 0$ while keeping the elastic constant $M\omega_0^2$ finite. In this regime, the phonon kinetic term $\hat{P}^2/2M$ becomes negligible. As the uniform shift $X_0$ simply renormalizes the chemical potential at half-filling, the resulting effective mean-field Hamiltonian  is
\begin{equation}
\mathcal{H}_{\rm MF}
=
-t_{1} \sum_{\langle i,j \rangle,\sigma}
( c^\dagger_{i\sigma} c_{j\sigma} + \mathrm{H.c.} )
+ t_{2} \sum_{\langle\!\langle i,j \rangle\!\rangle,\sigma}
( i\nu_{ij} \, c^\dagger_{i\sigma} c_{j\sigma} + \mathrm{H.c.} )
+ \sum_i \frac{M\omega_0^2}{2}\, \delta X ^2
- (\Delta + g\,\delta X)\sum_i \tau_i n_i.
\label{eq:MFT_hamiltonian_polished}
\tag{S2}
\end{equation}
This approximation yields a quadratic single-particle Hamiltonian that can be written in momentum space and diagonalized for each $\mathbf{k}$. For a given set of model parameters $(t_1, t_2, \Delta, g, M\omega_0^2)$, the sublattice distortion $\delta X$ is obtained by minimizing the Helmholtz free energy. For convenience, all mean-field results are expressed using the dimensionless electron-phonon coupling $\lambda = \frac{g^2}{M\omega_0^2}$. Figure~\ref{fig:mft} shows the zero-temperature results obtained from a numerical solution on a $100\times100$ momentum grid.

The essential mean-field trends agree with the finite-frequency QMC data. Panels (a) and (b) display an abrupt increase in $\delta X$ (and consequently in $\delta_n$) with increasing \textit{e–ph} coupling, indicating the absence of coexistence between the CDW and CI phases. When a sublattice potential is included, any finite value of $\Delta$ induces a finite charge imbalance, as shown in panels (c) and (d). With the mean-field solution for $\delta X$ in hand, the boundary of the CI phase is obtained exactly from the relation $\Delta + g\delta X = 3\sqrt{3}t_2$~\cite{haldane1988}, corresponding to the dashed line in panel (c). Although the numerical calculations were performed using $M\omega_0^2 = 1$, we note that the critical values of $\lambda$ are universal in the adiabatic limit and do not depend on the specific choice of $M\omega_0^2$, corresponding to the exact $\omega_0\to 0$ result. 

\section{Exact Diagonalization for the Haldane-Hubbard Model}
\label{supp:ED}

In the anti-adiabatic limit ($\omega_0 \to \infty$), the sign problem becomes severe, preventing us from accessing low temperatures and large system sizes within DQMC. In this regime, however, the Holstein coupling maps onto the attractive Hubbard interaction, enabling us to perform zero-temperature exact diagonalization for $L=3$. The simulations employ the Krylov-Schur eigensolver as implemented in the PETSc and SLEPc libraries~\cite{Balay,Slepc}.

\begin{figure}[H]
    \centering
    \includegraphics[scale= 0.5]{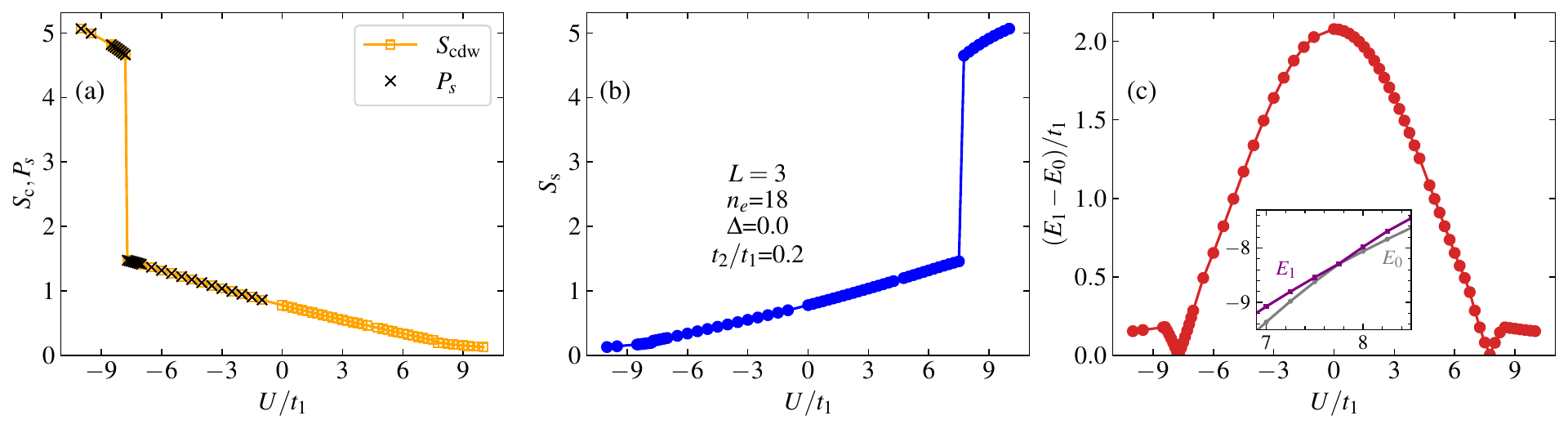}
    \caption{Exact diagonalization results: (a) charge ($S_c$) and $s$-wave pairing ($P_s$) structure factors, (b) antiferromagnetic structure factor ($S_s$), and (c) the energy gap between the ground state ($E_0$) and first excited state ($E_1$) as functions of the interaction strength $U/t_1$. The inset in panel (c) shows the energy-level crossing near the transition point $U/t_1 \approx 7.7$. All data correspond to a half-filled honeycomb lattice under periodic boundary conditions with fixed $L=3$, $\Delta=0.0$, and $t_2/t_1 = 0.2$.
    }
    \label{fig:supp_ED}
\end{figure}

In addition to the charge structure factor, we also probe superconductivity and magnetism through the $s$-wave pairing and antiferromagnetic spin structure factors,
\begin{equation}
    P_s = \frac{1}{N} \sum_{\iv,\jv,\s} \langle c^\dagger_{\iv,\s}c^\dagger_{\iv,\sp} c_{\jv,\sp}c_{\jv,\s} \rangle ,
    \quad \text{ and } \quad
    S_{\text{s}}\! =\! \frac{1}{N} \sum_{\rv_{\iv}, \rv_{\jv}} \!\langle(S^z_{A,\rv_{\iv}}\!\! -\! S^z_{B,\rv_{\iv}})(S^z_{A,\rv_{\jv}}\!\! - \!S^z_{B,\rv_{\jv}}) \rangle.
\label{eq:ssdw}
\tag{S3}
\end{equation}
Combining these quantities with the energy gap $\Delta E = E_1 - E_0$, we identify a first-order transition separating the CI phase from the CDW + SC state at $U/t_1 \approx -7.7$, as shown in Fig.~\ref{fig:supp_ED}(a). For $U/t_1 > 0$, the spin structure factor $S_s$ mirrors the behavior of $S_c$ across $U=0$, and the critical magnitude $|U_c/t_1| \approx 7.7$ remains the same, as illustrated in Fig.~\ref{fig:supp_ED}(b). This correspondence is expected: a particle-hole transformation maps the attractive and repulsive Hubbard models when the next-nearest-neighbor hopping is purely imaginary~\cite{Zheng2011}. Nevertheless, this gauge transformation does not eliminate the sign problem, unlike in the Kane-Mele–Hubbard model~\cite{Hohenadler2011,Zheng2011}.

This mapping also allows comparison with DMRG results for the repulsive case in Ref.~\cite{He2024}, which finds a critical value $U/t_1 \approx 6.8$ in infinite cylinder geometries. The energy spectrum further confirms the first-order character of the transition: an energy-level crossing occurs at the critical point, accompanied by a closing of the many-body gap, as shown in Fig.~\ref{fig:supp_ED}(c).

\section{Sublattice resolved Green's function}
\label{supp:Gktau}

The complete momentum-dependent unequal time  Green's function can be written as
\begin{equation}
G(\mathbf{k},\tau)
=
\begin{pmatrix}
G_{AA}(\mathbf{k},\tau) & G_{AB}(\mathbf{k},\tau) \\
G_{BA}(\mathbf{k},\tau) & G_{BB}(\mathbf{k},\tau)
\end{pmatrix}.
\tag{S4}
\end{equation}
where we define $G_{AA}(\mathbf{k},\tau) \equiv G_{A}(\mathbf{k},\tau)$ and $G_{BB}(\mathbf{k},\tau) \equiv G_{B}(\mathbf{k},\tau)$. With this,
we take the trace of the Green's function to construct the total diagonal contribution, 
\begin{equation}
\mathrm{Tr}\!\left[ G(\mathbf{k},\tau) \right]
=
G_{AA}(\mathbf{k},\tau) + G_{BB}(\mathbf{k},\tau) \equiv G_{\rm {total}} (\mathbf{k}, \tau).
\tag{S5}
\end{equation}

\end{document}